\documentclass[prl,twocolumn,superscriptaddress,showpacs]{revtex4-1}
\usepackage{graphicx}
\usepackage{amsmath}
\let\oldsqrt\sqrt
\def\sqrt{\mathpalette\DHLhksqrt}
\def\DHLhksqrt#1#2{%
\setbox0=\hbox{$#1\oldsqrt{#2\,}$}\dimen0=\ht0
\advance\dimen0-0.2\ht0
\setbox2=\hbox{\vrule height\ht0 depth -\dimen0}%
{\box0\lower0.4pt\box2}}

\usepackage{color}

\definecolor{cola}{rgb}{0.7,0.1,0.1}
\definecolor{colb}{rgb}{0.9,0.4,0}
\definecolor{colc}{rgb}{0.3,0.7,0}

\begin{document}

%title and top stuff
\title{Efficient extraction of zero-phonon-line photons from single nitrogen-vacancy centers in an integrated GaP-on-diamond platform}
\author{Michael Gould} 
\email{gouldm2@uw.edu}
\affiliation{Department of Electrical Engineering, University of Washington, Seattle WA 98195 USA}
\author{Emma R. Schmidgall}
\affiliation{Department of Physics, University of Washington, Seattle WA 98195 USA}
\author{Shabnam Dadgostar}
\affiliation{Department of Physics, Humboldt-Universitat zu Berlin, Newtonstrasse 15, 12489 Berlin, Germany}
\author{Fariba Hatami}
\affiliation{Department of Physics, Humboldt-Universitat zu Berlin, Newtonstrasse 15, 12489 Berlin, Germany}
\author{Kai-Mei C. Fu}
\affiliation{Department of Electrical Engineering, University of Washington, Seattle WA 98195 USA}
\affiliation{Department of Physics, University of Washington, Seattle WA 98195 USA}

\begin{abstract} 

Scaling beyond two-node quantum networks using nitrogen vacancy (NV) centers in diamond is limited by the low probability of collecting zero phonon line (ZPL) photons from single centers. Here, we demonstrate GaP-on-diamond disk resonators which resonantly couple ZPL photons from single NV centers to single-mode waveguides. In these devices, the probability of a single NV center emitting a ZPL photon into the guided waveguide mode after optical excitation can reach 9\%, due to a combination of resonant enhancement of the ZPL emission and efficient coupling between the resonator and waveguide. We verify the single-photon nature of the emission and experimentally demonstrate both high in-waveguide photon numbers and substantial Purcell enhancement for a set of devices. These devices may enable scalable integrated quantum networks based on NV centers. 
\end{abstract}

\maketitle

The negatively charged nitrogen-vacancy (NV) center in diamond shows significant promise as a solid-state qubit register \cite{ref:gurudevdutt2007qrb,ref:waldherr2014qec,ref:blok2015tqn} for measurement-based quantum information processing (MBQIP)\cite{ref:raussendorf2001aow, ref:benjamin2006bgs, ref:li2012htd}. The computational resource in MBQIP is a network of entangled qubit registers.  For NV centers, this network can be grown via single photon measurement of the NV zero-phonon line (ZPL) emission~\cite{ref:cabrillo1999ces,ref:barrett2005ehf}.  Two-qubit networks of NV centers have been heralded in this manner using free-space collection optics \cite{ref:bernien2013heb, ref:Pfaff2014uqt}. However, the demonstrated entanglement generation rate was significantly slower than the electron spin decoherence rate, and thus far too slow to allow multi-qubit entanglement. The limiting factor in reported entanglement rates is the low probability of detecting a ZPL photon upon excitation of an NV center. We will call this probability the total quantum efficiency, $\eta$. Successful entanglement is heralded by two independent ZPL photon detection events~\cite{ref:barrett2005ssn}, and thus the entanglement generation rate scales as $\eta^2$. Low achieved $\eta$ values are primarily the result of two effects inherent to NV centers in diamond. First, the high refractive index of diamond limits free-space collection efficiency through total internal reflection. Second, phonon interactions result in only $\sim$3\% of radiative emission occurring via the ZPL transition~\cite{ref:barclay2011hnr, ref:davies1974vsd, ref:siyushev2009lto}. Photonic device integration can mitigate both effects, providing a scalable photonics platform for building quantum networks.

In this work we demonstrate a key step toward realizing such a network in a GaP-on-diamond integrated photonics platform: the efficient optical coupling of single NV centers to single-mode waveguides. We show that the probability of emitting a ZPL photon into the guided mode after optical excitation can reach 9\%. This high probability is achieved through a combination of resonant enhancement of ZPL emission via the Purcell effect~\cite{ref:purcell1946sep}, as well as efficient coupling between the resonant devices and waveguides. 10 out of 80 tested devices exhibited resonantly enhanced waveguide collection rates which exceed the theoretical limit for non-resonant collection. The limiting factor for yield is the NV-cavity coupling which can be readily improved with NV-cavity registration. Furthermore, the devices were fabricated on the same chip as passive integrated photonic components~\cite{ref:gould2016lsg} necessary for on-chip entanglement generation networks. Combined, these results indicate the promise of the GaP-on-diamond photonics platform for scalable quantum networks.

Our platform utilizes a 125~nm thick GaP membrane to guide optical modes at the surface of the diamond chip~\cite{ref:barclay2011hnr, ref:gould2016lsg}, taking advantage of the high refractive index of GaP (n = 3.3) compared to that of diamond (n = 2.4).  This is in contrast to the more common approach utilizing the diamond itself as the waveguiding material~\cite{ref:loncar2013qpn}. A key advantage of the hybrid platform is fabrication scalability. Specifically, diamond waveguides require either undercutting of the diamond~ \cite{ref:burek2014hqf,ref:khanaliloo2015hqv} or working with thin diamond membranes on a low index substrate~\cite{ref:faraon2013qpd, ref:hausmann2012idn}. Undercutting requires a three-dimensional dry etch, significantly constraining the device layout. Thin diamond membranes with large area and uniformity have yet to be demonstrated, resulting in poor device uniformity across a chip. On the other hand, large-area (cm-scale) highly uniform GaP membranes can be grown epitaxially and transferred to bulk single-crystal diamond chips, enabling the fabrication of large numbers of photonic devices with good cross-chip uniformity. For complex photonic circuits, additional features of the GaP-on-diamond platform include the introduction of a second-order optical non-linearity~\cite{ref:nelson1968eop}, which should enable active photonic routing, and sub-nm top-surface roughness suitable for the development of on-chip superconducting nanowire single-photon detectors~\cite{Sprengers2011,Akhlaghi2015,Pernice2012}. The primary disadvantage associated with the hybrid-material platform is the inherently weaker coupling between the emitters located in the diamond and the guided optical modes primarily localized in the GaP. However, as we demonstrate below, this effect is mitigated with resonant devices of sufficiently high quality factor. 

\begin{figure}
    \includegraphics[width=0.45\textwidth]{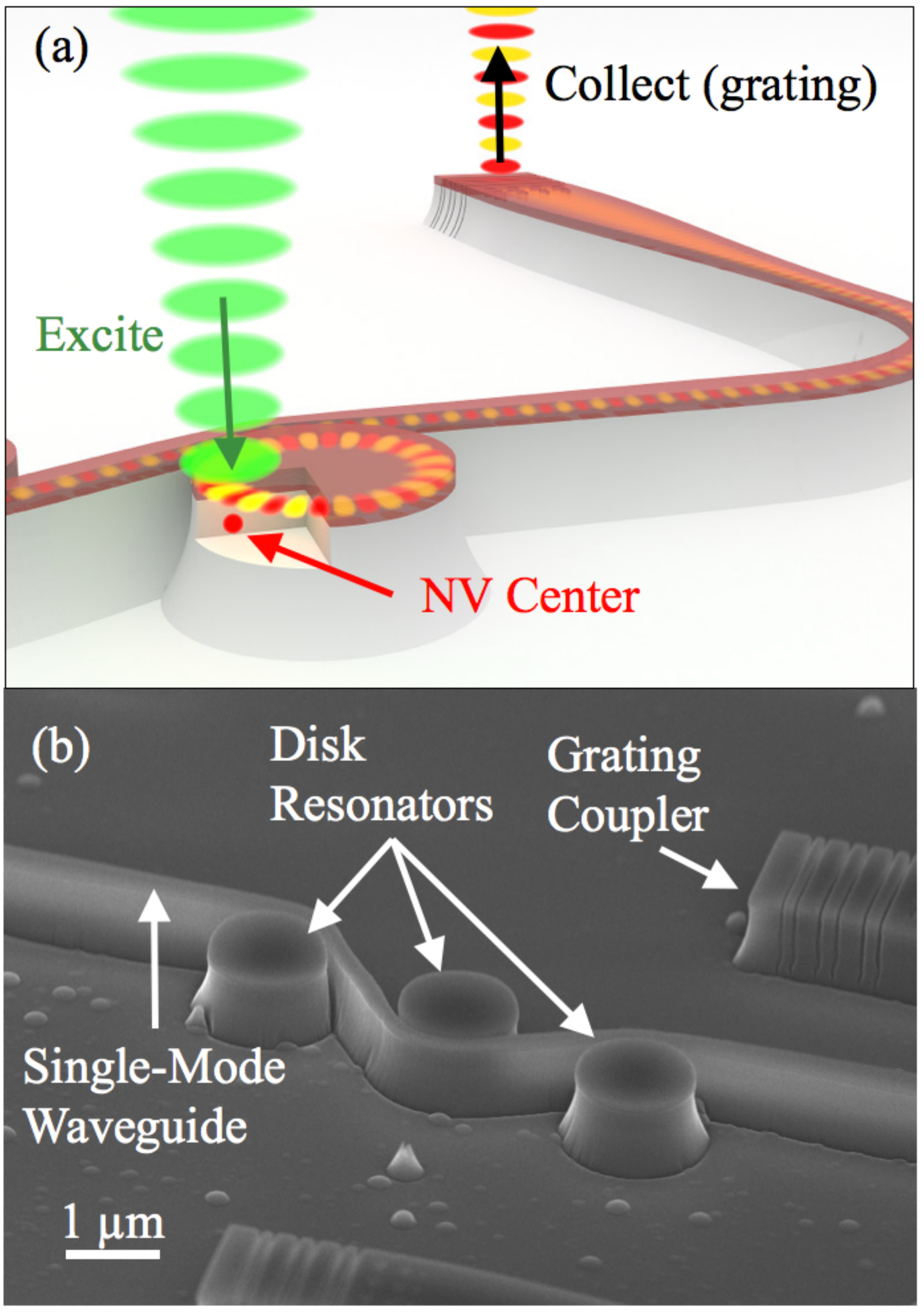}
  \caption{(a) Illustration of device measurement showing grating-collection. (b) Scanning electron microscope image of fabricated devices.}
  \label{fig:schematic}
\end{figure}

The photon-collection devices consist of near-surface NV centers evanescently coupled to the fundamental TE-polarized whispering-gallery mode of 1.3~$\mu$m-diameter disk resonators. The resonators are coupled to 150~nm-wide single-mode ridge waveguides (Fig.~\ref{fig:schematic}a,b).  Prior to device fabrication, near-surface NV centers were created in the single-crystal electronic-grade diamond chip by nitrogen ion-implantation and annealing. A 125-nm GaP membrane was then transferred onto the diamond via epitaxial lift-off and van der Waals bonding~\cite{ref:yablonovitch1990vdw}. Devices were fabricated on the resulting GaP-on-diamond chip by electron-beam lithography and reactive ion etching~\cite{ref:thomas2014wis}. The resulting device cross-section is a 125 nm GaP waveguiding layer on a ~600 nm diamond pedestal, with a sparse layer  of NV centers ($\sim2\times10^9$~cm$^{-2}$) in the top 10-20 nm of the diamond. Further fabrication details are given in the Supplemental Information~\cite{ref:SI}.

\begin{figure}
\begin{center}
\includegraphics[width=0.45\textwidth]{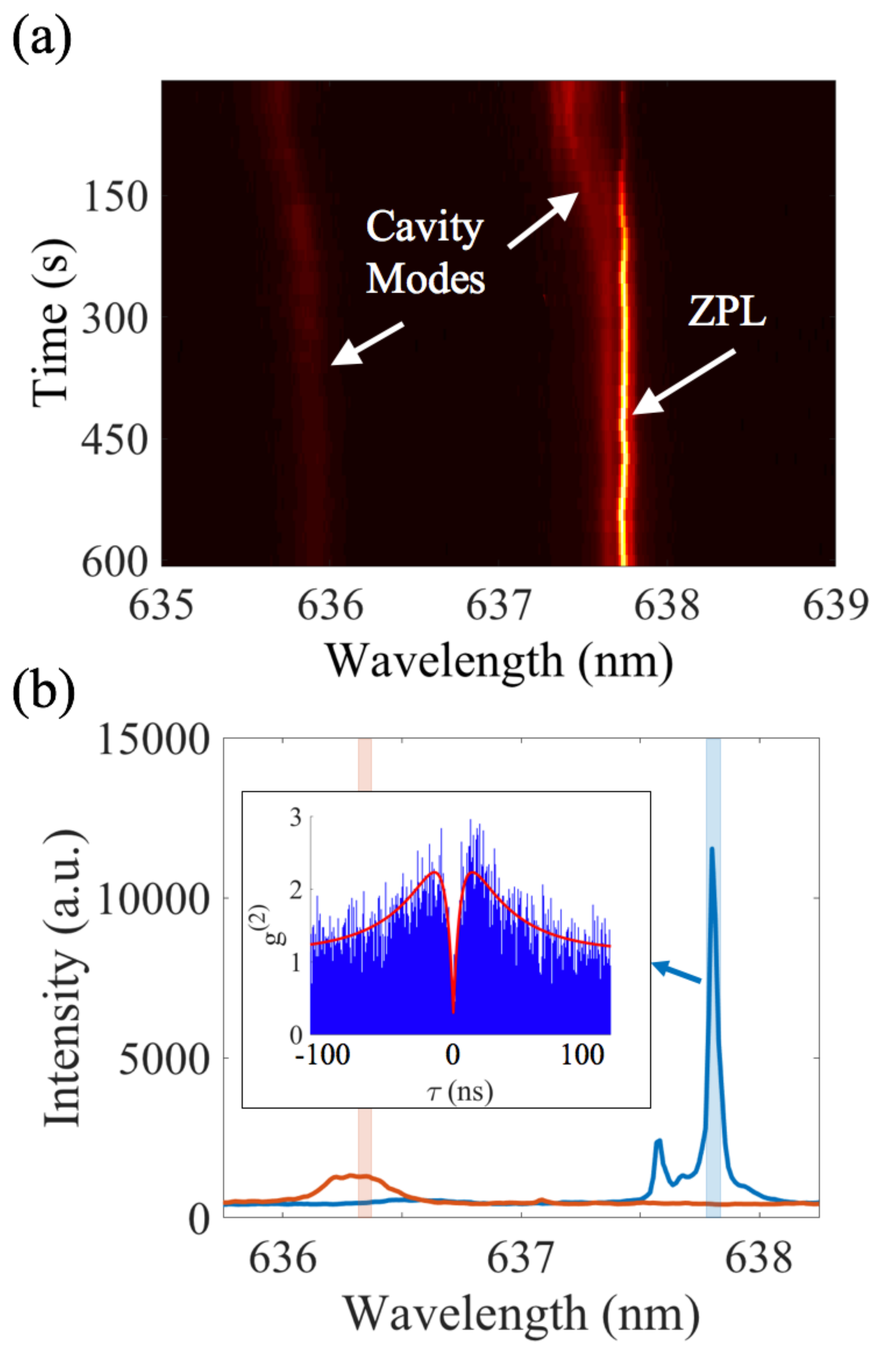}
\end{center}
\caption{(a) Measured tuning curve showing two cavity modes as one is tuned onto resonance with a coupled NV center's ZPL. (b) Grating-collected spectra with cavity tuned onto selected ZPL (blue curve) and detuned from ZPL (orange curve), with wavelength integration range used for count rate calculation indicated by shaded rectangles. Inset: photon autocorrelation measured with cavity tuned onto ZPL, with bi-exponential fit shown in red.}
\label{fig:spectra}
\end{figure}

Measurements were performed with the fabricated devices cooled to 8~K. For each device, the resonator mode was first tuned to the ZPL resonance. Tuning was accomplished via xenon gas deposition, which causes the cavity modes to red-shift and provides a wavelength tuning range of $\sim$2~nm. For cavity tuning measurements, the sample was excited at normal incidence and fluorescence spectra were collected from the output grating coupler as illustrated in Fig.~\ref{fig:schematic}a.  An example tuning curve which shows clear NV-cavity coupling as the cavity is tuned to the NV ZPL resonance is shown in Fig.~\ref{fig:spectra}a. This initial tuning measurement was performed on approximately 80 devices expected to lie within the cavity tuning range of the ZPL wavelength, for four different excitation locations around the perimeter of each disk. In this way, a subset of devices showing coupled ZPL emission were identified for further study. Three additional types of measurement were performed on devices in this subset: photon auto-correlation ($g^{(2)}$) on the grating-coupled ZPL emission to confirm the single-photon nature of the collected fluorescence, power dependence to determine saturated collection rates, and lifetime measurements to quantify the resonant enhancement of the ZPL emission. As we show below, the last two measurements enable two separate estimates of $\eta$ for each device.

With a cavity mode tuned onto resonance with a selected ZPL, we first performed a $g^{(2)}$ measurement to verify the single-emitter nature of the source (see inset Fig.~\ref{fig:spectra}b).  The grating-collected light was spectrally filtered around the selected ZPL wavelength before detection as depicted in Fig.~\ref{fig:spectra}b. The  $g^{(2)}$ measurement was performed on 4 of the brightest devices, all showing auto-correlation dips with $g^{(2)}(0) < 0.4$, indicating that in each device a majority of the collected photons are from a single emitter. Non-zero coincidence rates are the result of background fluorescence at the ZPL wavelength. This background fluorescence can be observed in the detuned-cavity spectrum (orange) in Fig.~\ref{fig:spectra}b.

We next measured the excitation power dependence of the waveguide-coupled ZPL photon rate to determine saturated collection rates. This measurement was performed by sweeping the excitation power and measuring the grating-coupled detection rate, again spectrally filtered around the selected ZPL. After removal of the background fluorescence, measured with the cavity mode detuned from the ZPL, the data were fit to a saturation model: $\gamma(P) = \gamma_{sat}/(1 +P/P_{sat})$, where $\gamma(P)$ is the detection rate, $\gamma_{sat}$ is the saturated detection rate, $P$ is the excitation power and $P_{sat}$ is the saturation power. 

\begin{figure}
\begin{center}
\includegraphics[width=0.4\textwidth]{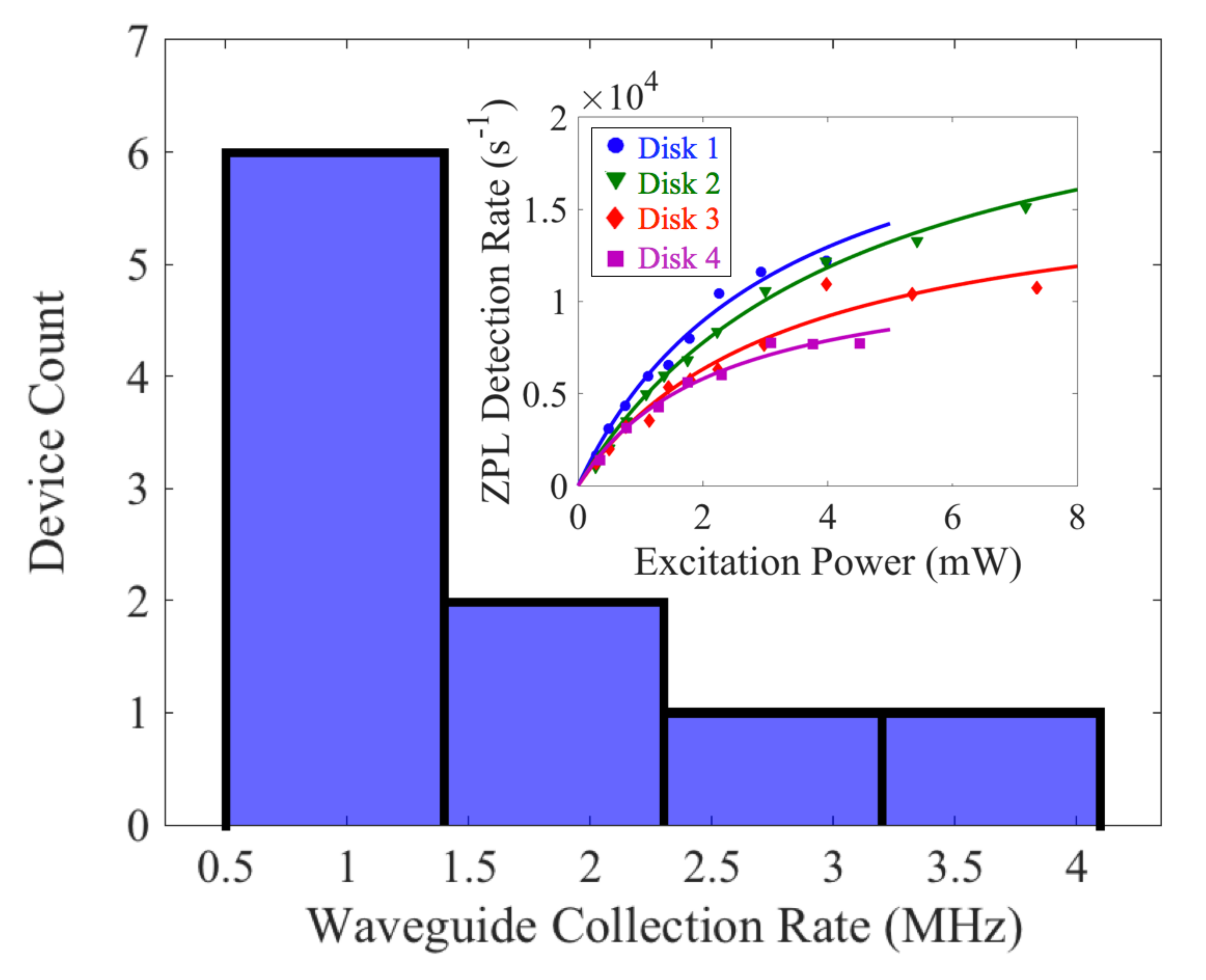}
\end{center}
\caption{Histogram of estimated saturated collection rates into bus waveguides for 10 devices. Inset: power dependence of on-resonance NV ZPL detection rate for 4 selected devices, with background removed. Solid lines are fits to saturation model.}
\label{fig:counts}
\end{figure}

Power dependence data for 4 devices are shown in Fig.~\ref{fig:counts} (inset). Disk 1 shows a detected ZPL count rate of 1.2$\times10^4$~s$^{-1}$ after background subtraction. The fit indicates a saturated NV ZPL detection rate of 2.0$\times10^{4}$~s$^{-1}$, and a saturation power of 3.4 mW. Using the measured collection path efficiency for each device~\cite{ref:SI} and the known detector efficiency, we can estimate the saturated collection rate into the bus waveguide. In the case of Disk 1, the estimated on-chip collection rate is 2.5$\times10^6$~s$^{-1}$ from a single saturated NV center. Figure~\ref{fig:counts} shows a histogram of saturated on-chip collection rates for 10 devices with values exceeding 5$\times10^5$~s$^{-1}$. We note that the estimated count rates for the three brightest devices are comparable to the best reported collection rates of NV ZPL photons into guided modes for all-diamond devices~\cite{ref:faraon2013qpd}. These count rates are also several times larger than the theoretical limit of approximately $3\times10^5$~s$^{-1}$ in the absence of Purcell enhancement, calculated as 3\% of a total saturated emission rate of $1\times10^7$~s$^{-1}$~\cite{Jelezko2006}.

\begin{figure}
\begin{center}
\includegraphics[width=0.45\textwidth]{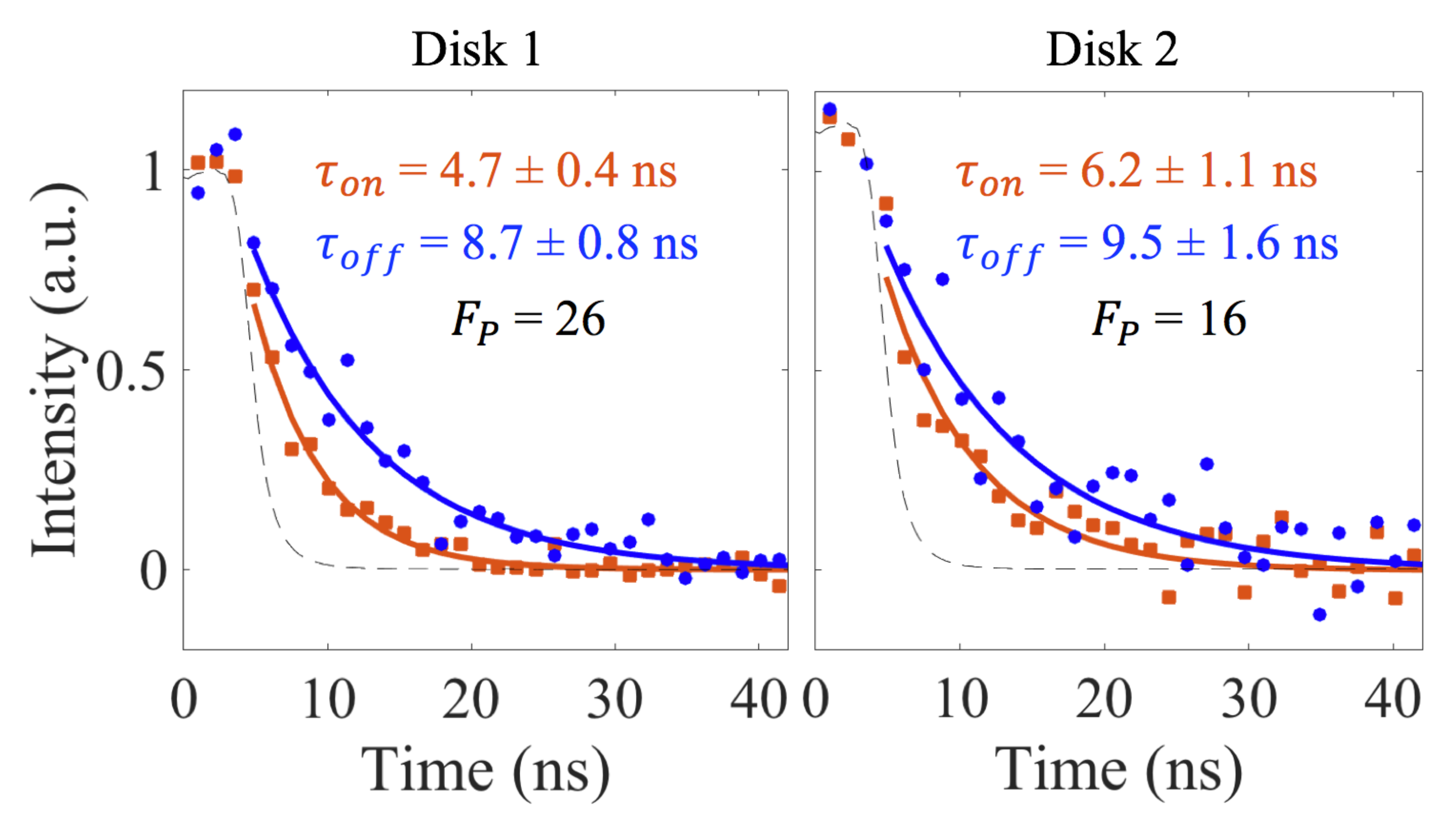}
\end{center}
\caption{Fluorescence lifetime measurements for Disks 1 and 2, in both the resonant (orange) and off-resonant (blue) conditions. Thick lines represent exponential fits. The dashed black line is the measured system response for reflected excitation light.}
\label{fig:lifetimes}
\end{figure}

In order to quantify the achieved Purcell enhancement, the excited-state lifetimes of individual NV centers were measured using a directly modulated laser diode~\cite{ref:oeckinghaus2014cdl} with a measured fall~time of 1 ns. Time-resolved measurements were taken on Disks 1-4, with the cavity on-resonance with a selected ZPL, as well as off-resonance. After careful subtraction of the background fluorescence waveform~\cite{ref:SI}, the data were fit to exponential decay curves to obtain the lifetimes. Fig.~\ref{fig:lifetimes} depicts measured on- and off-resonance time-resolved photoluminescence curves for Disks 1 and 2. Measured lifetimes under both resonance conditions were compared in order to determine the Purcell enhancement factor $F_P$ of a given device.  In the non-resonant case, the lifetime $\tau_0$ is determined by $1/\tau_0 =\Gamma_0 =  \Gamma_{ZPL} + \Gamma_{PSB}$,  in which $\Gamma_{ZPL}$ ($\Gamma_{PSB}$) is the emission rate into the ZPL (phonon sidebands). In the resonant case, the lifetime $\tau_{res}$ is determined by $1/\tau_{res} = \Gamma_{res} = (1 + F_P)\Gamma_{ZPL} + \Gamma_{PSB}$. For Disk 1, the measured on-resonance lifetime of 4.7 $\pm$ 0.4 ns is significantly shorter than the off-resonance lifetime of 8.7 $\pm$ 0.8 ns, with the ratio corresponding to a resonant Purcell factor of $F_P$ = 26. This is close to the maximum possible $F_{P,max}$ $\approx$ 30 for this device geometry, given a measured quality factor of Q = 8200 ~\cite{ref:thomas2014wis}.  We note that the off-resonance lifetimes in all four measured devices are significantly shorter than the NV lifetime in bulk diamond ($\sim$12 ns) \cite{Batalov2008,Collins1983}. The shorter lifetimes are consistent with a broadband enhancement effect caused by the NV centers' proximity to the diamond-GaP interface~\cite{ref:SI}. 

The total quantum efficiency $\eta$  was estimated for Disks 1-4 by two different methods. In the first method, the measured on-resonance lifetime was used to estimate the total saturated emission rate, $\gamma_{tot}$ of a selected NV center~\cite{ref:SI}. For a saturated on-chip collection rate of $\gamma_{wg}$, $\eta_1$ is given by
\begin{equation}
	\eta_{1} = \frac{\gamma_{wg}}{\gamma_{tot}}.
\end{equation}
In the case of Disk 1, $\eta_{1} \approx 9\%$. The second method uses the total quantum efficiency into the disk resonator mode, $\eta_{disk}$, calculated from the measured Purcell enhancement factor; and the disk-to-waveguide out-coupling efficiency, determined from grating-coupled transmission measurements~\cite{ref:SI}. In this case, 
\begin{equation}
	\eta_{2} = \eta_{out}\eta_{disk} = \frac{\eta_{out}F_P}{F_P + \Gamma_0/\Gamma_{ZPL}},
\end{equation}
where $\eta_{out}$ is the disk-to-waveguide out-coupling efficiency and $\Gamma_0/\Gamma_{ZPL} \approx 30$. For Disk 1, $\eta_2\approx 9\%$. Table~\ref{table} summarizes the estimated total quantum efficiency obtained using both methods for 4 devices, showing reasonable agreement between the two.

\begin{table}[h!]
\begin{center}
	\begin{tabular}{|c || c | c | c | c | c | c | c |}
		 \hline		
			 \textbf{Device} & $\mathbf{\gamma_{tot} (s^{-1})}$ & $\mathbf{\gamma_{wg} (s^{-1})}$ & $\mathbf{\eta_{1}}$ & $F_P$ & $\mathbf{\eta_{out}}$ &  $\mathbf{\eta_{2}}$ & $g^{(2)}(0)$\\ \hline \hline
			 Disk 1 & $2.85 \times 10^7$ & $2.48 \times 10^6$ & $9\%$ & $26$ & $20\%$ & $9\%$ & $0.30$\\ \hline
			 Disk 2 & $2.15 \times 10^7$ & $2.17 \times 10^6$ & $10\%$ & $16$ & $23\%$ & $8\%$ & $0.36$\\ \hline
			 Disk 3 & $1.85 \times10^7$ & $1.48 \times 10^6$ & $8\%$ & $12$ & $12\%$ & $3\%$ & $0.19$\\ \hline
			 Disk 4 & $2.49 \times10^7$ & $9.72 \times 10^5$ & $4\%$ & $16$ & $12\%$ & $4\%$ & $0.31$\\ \hline	
	\end{tabular}	
\end{center}
	\caption{Summary of key values for 4 selected devices.}
	\label{table}
	\end{table}

We have shown that large $\eta$ values are achievable in a GaP-on-diamond platform, using devices that can be readily integrated into larger on-chip photonic networks. A reasonable excitation repetition rate for NV-NV entanglement is 100~kHz, limited by the NV initialization  time~\cite{ref:Pfaff2014uqt}. If all waveguide-coupled photons are detected and indistinguishable, the demonstrated collection efficiency of 9\% would correspond to an NV-NV entanglement generation rate of 400 Hz, which significantly exceeds the $\sim$1 s electron spin decoherence rate~\cite{ref:bargill2013sse}. We note that the GaP-on-diamond system is compatible with waveguide-coupled superconducting detectors, a technology which has already demonstrated detection efficiencies exceeding 90\% for waveguide-coupled photons~\cite{Akhlaghi2015, Pernice2012}.

The demonstrated Purcell factors, as high as 26, exceed what has been achieved in all-diamond waveguide-integrated platforms~\cite{ref:faraon2011rez,ref:faraon2013qpd}. This suggests that the primary disadvantage of the hybrid materials system for MBQIP, namely that the emitter cannot be placed at the guided-mode maximum, can be largely overcome with continued improvements in resonator quality factor. A greater challenge for all integrated platforms is the production of indistinguishable photons. Specifically, it will be necessary to improve the spectral stability of near-surface NV centers, which currently exhibit spectral diffusion up to 10~GHz~\cite{ref:fu2010cnn}.  We are encouraged by recent work in improving NV spectral stability via high-temperature annealing~\cite{ref:chu2014cot} and longer-wavelength excitation~\cite{ref:Pfaff2014uqt}. Moreover, even if device-integrated NV centers do not exhibit the spectral stability observed for bulk NV centers incorporated during diamond growth, the platform is compatible with Stark tuning for both active ZPL frequency stabilization~\cite{ref:acosta2012dso} and tuning to a single platform resonance~\cite{ref:bernien2013heb,ref:bassett2011ets}.

We conclude with an outlook for scalability.  Our yield for simple photonic circuits which couple the ZPL emission from a single NV center to a single-mode waveguide, and which outperform the theoretical limit for free-space collection, exceeds 10\%. This yield is predominantly limited by the yield in NV-resonator coupling, which in the short-term can be improved by increasing the density of near-surface NV centers. Longer-term, aligned implantation~\cite{ref:toyli2010csn}, combined with on-chip switching for device post-selection, should enable deterministic coupling of high-performing devices. For the latter approach, the second-order optical non-linearity associated with GaP can be leveraged to implement integrated electro-optic switching. Thus we believe the high total quantum efficiency $\eta$, combined with large-scale integration demonstrated in this work, is a promising step toward quantum photonic networks in the hybrid GaP-on-diamond platform.

{\it Acknowledgements:} This material is based upon work supported by the National Science Foundation under Grant Number 1506473.
We would like to acknowledge I.R.Christen for assistance with graphics, I.R. Christen, S.Chakravarthi for passive device testing, Y.Zhou for FDTD simulations of fabricated devices, and R. Bojko for e-beam lithography support. Devices were fabricated at the Washington Nanofabrication Facility, a part of the National Nanotechnology Coordinated Infrastructure network.

\clearpage
\part{Supplemental Information}

%title and top stuff

\section{Device fabrication}

Near-surface NV centers were created in a single-crystal, electronic-grade diamond chip (ElementSix) by N$^+$ ion implantation (10~keV, $1\times10^{10}$ cm$^{-2}$, CuttingEdge Ions), followed by a two-step anneal. A 1-hour, $850^{\circ}$ C annealing step was performed under a 5\%/95\% $\mathrm{H}_{2}/\mathrm{Ar}$ forming gas atmosphere in order to allow diffusion of vacancies to form NV centers. A subsequent 24-hour, $450^{\circ}$ C anneal was performed in air in order to oxygen-terminate the surface, a necessary step to ensure stability of the negatively charged state of near-surface NV centers~\cite{Fu2010}.

Following the epitaxial lift-off and transfer of the GaP membrane to the implanted diamond chip, hydrogen silsesquioxane (HSQ) was spin-coated to be used as an electron-beam lithography resist. Devices were patterned by electron-beam lithography (JEOL 6300), followed by two reactive-ion etch (RIE) steps. The first RIE step (3.0 mTorr, 1.0/6.0/3.0 sccm $\mathrm{Cl}_2/\mathrm{Ar}/\mathrm{N}_2$) was used to etch through the GaP, and the second (25.0 mTorr, 20 sccm $\mathrm{O}_2$) was used to etch into the diamond.

\section {Collection Path Efficiency}

In order to obtain accurate estimates of waveguide collection rates, the collection path efficiency ($\eta_{cp}$) from the coupled section of the waveguide to the off-chip detector was measured for each device. The collection path can be separated into 2 main components: the chip and the microscope. The microscope collection path efficiency ($\eta_{mic}$) is assumed to be the same for all devices, and was measured by passing a laser beam ($\lambda\approx$ 640 nm) through the system, and measuring the input and output power. The measured efficiency of the microscope was approximately $\eta_{mic}$ = 35\%, with the majority of the losses coming from the grating spectrometer used for spectral filtering of the ZPL ($\eta_{spec}$ = 45\%).

\begin{figure*}
  \begin{center}
    \includegraphics[width=15cm]{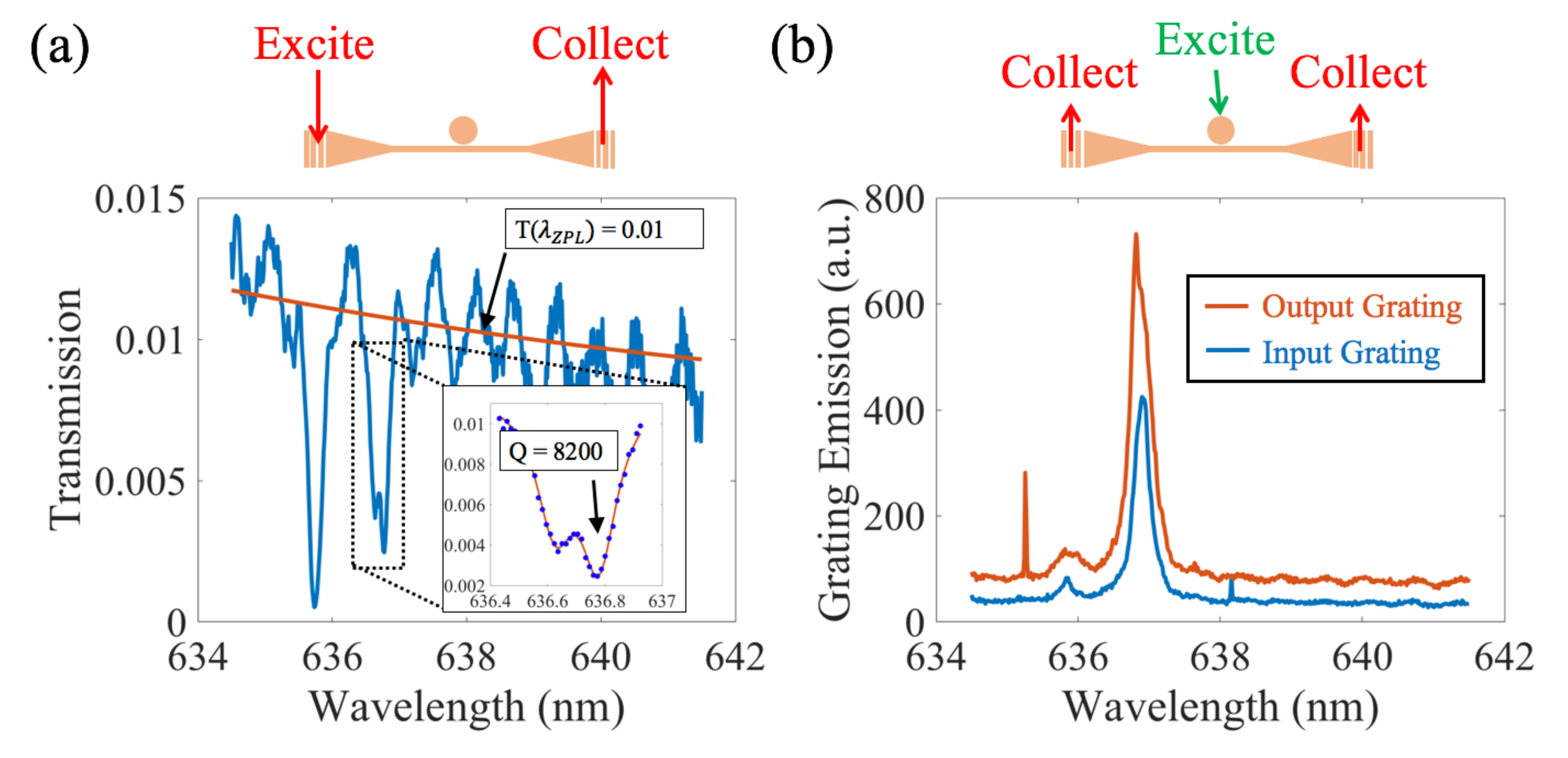}
  \end{center}
  \caption{(a) Measured transmission spectrum for Disk 1 showing Mie splitting, a quality factor of Q = 8200, and a transmission value of 0.01 at the ZPL wavelength ($\lambda_{ZPL}$). A shorter-wavelength resonance dip visible at $\lambda$ = 635.8 nm is due to another disk coupled to the same waveguide. (b) Emission spectra from input and output grating couplers, showing an output efficiency ratio R = 1.65. Top schematics: measurement setup in each case.}
\label{fig:transmission/emission}
\end{figure*} 

The chip efficiency ($\eta_{chip}$) was determined for each device individually. First, grating-coupled transmission measurements were performed and the results were normalized to power reflected from a polished diamond surface (Fig.~~\ref{fig:transmission/emission}a). In order to account for differences between the `input-to-device' ($\eta'_{chip}$) and `device-to-output' ($\eta_{chip}$) efficiencies, emission was measured from each grating coupler while exciting background fluorescence in the selected device (Fig.~\ref{fig:transmission/emission}b). Using these two measurements, the chip efficiency (assuming collection through the output grating coupler) was determined as:

\begin{equation}
	\eta_{chip} = \sqrt{TR},
\end{equation}
where T = $\eta_{chip}\eta'_{chip}$ is the measured transmission, and R = $\eta_{chip}/\eta'_{chip}$ is the ratio of background fluorescence intensity from the output/input grating couplers. In the case of Disk 1, the transmission at the ZPL wavelength ($\lambda_{ZPL}$) was determined to be T = 0.01, and the output-to-input ratio was R = 1.65, yielding an on-chip efficiency of $\eta_{chip} = 12.8\%$.

Combining the measured chip and microscope efficiencies, as well as the known spectrometer CCD detection efficiency of $\eta_{det} = 0.4$, we determine the total optical loss in the measurement. The total collection-path efficiency for a photon collected into the waveguide is given by:

\begin{equation}
	\eta_{cp} = \frac{1}{2}\eta_{chip}\eta_{mic}\eta_{det}
\end{equation}
where the factor of $1/2$ is included to account for the equal number of photons collected in the waveguide's counter-propagating mode. This assumes optical reciprocity and the electric dipole nature of NV-center emission. The total collection-path efficiency for Disk 1 calculated in this way is $\eta_{cp} = 0.9\%$.

\section{Lifetime Measurements }

A directly driven laser-diode (PicoLas, Roithner) emitting at a wavelength of 520 nm was used for lifetime measurements~\cite{Oeckinghaus2014}. This was the longest wavelength laser diode commercially available to us below 600 nm. Unfortunately, the photon energy is larger than the GaP bandgap, resulting in a large increase in background fluorescence relative to NV center ZPL emission. For this reason, careful background removal was necessary in order to obtain accurate lifetime data for coupled NV centers.

\begin{figure*}
  \begin{center}
    \includegraphics[width=15cm]{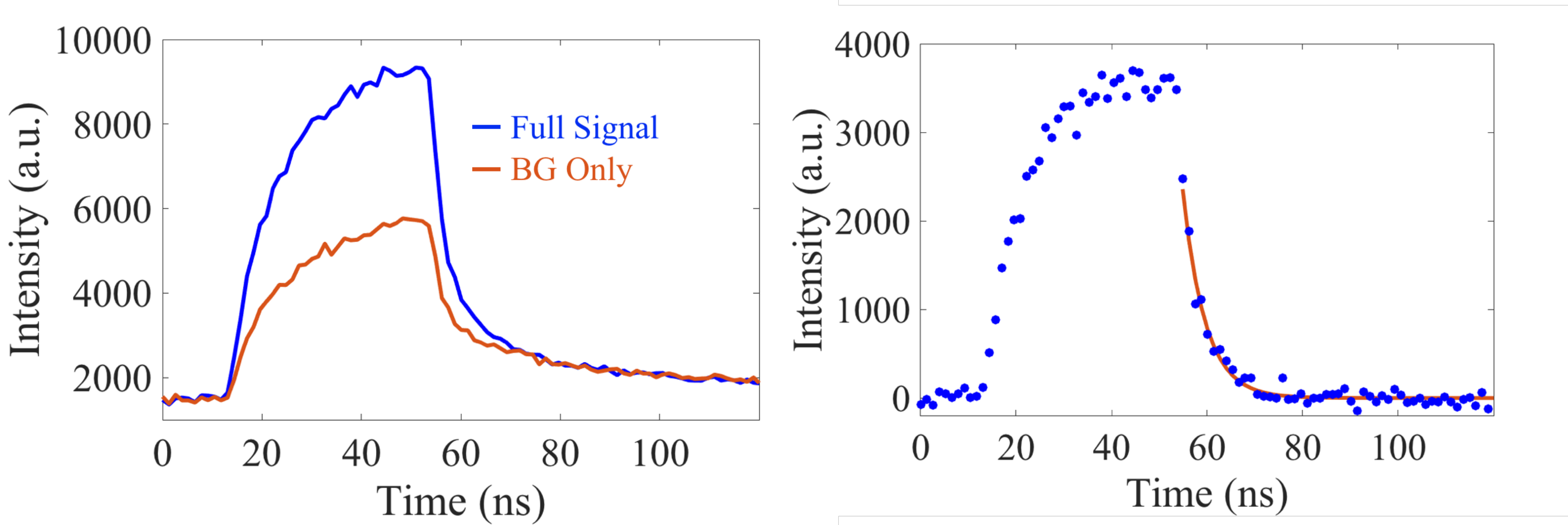}
  \end{center}
  \caption{(a) Full signal time-resolved fluorescence data for Disk 1 (blue curve) and background-only data (orange curve). (b) Time-resolved fluorescence after background removal, with a single exponential fit (orange curve).}
\label{fig:lifetime}
\end{figure*}

Both full-signal and background-only time-resolved photoluminescence measurements were taken for each resonant state, for each device (Count Blue single-photon counting module, ID Quantique ID801). The background was then scaled to match the full-signal at times much longer than the NV lifetimes, as depicted in Fig.~\ref{fig:lifetime}a. For the on-resonance case, background-only data was taken as the cavity-mode fluorescence with the cavity detuned from the NV center ZPL, while exciting at the NV center's location. The raw full-signal and scaled background histograms for Disk 1 (on-resonance, grating-collected) are shown in Fig.~\ref{fig:lifetime}a. Lifetimes were obtained using weighted exponential fits~\cite{Turton2003} after background removal (Fig.~\ref{fig:lifetime}b).

The grating-collected cavity-mode background fluorescence showed two distinct lifetime scales, one on the order of a nanosecond (possibly measurement-limited) and a second on the order of $\sim$30 ns. This longer-lived background fluorescence is greatly reduced when exciting on other parts of the disk off of the NV center location, or when collecting at a wavelength away from the cavity mode resonance. This indicates the coupled NV center is the source of this long-lived fluorescence, which we tentatively attribute to device-coupled phonon-sideband emission associated with the NV center's neutral charge state~\cite{Liaugaudas2012}. The faster decaying component is likely the result of recombination via deep levels in the GaP or fluorescence from the HSQ electron-beam resist.

We also note that the off-resonance lifetimes for all four measured devices are significantly shorter than what is typically reported for NV centers in bulk diamond ($\sim$12 ns). This is consistent with a model for emitters in close proximity to an infinite 2D dielectric interface~\cite{Lukosz1977}. In the case of NV centers at distances of 0 to 20 nm from a diamond-GaP interface along a [100] plane, the model predicts reduced lifetimes from 8 ns to 10 ns.

\section{Total Quantum Efficiency Estimation}

Total quantum efficiency ($\eta$) values were estimated using two methods. In the first method, $\eta_1$ is determined as the ratio of the saturated on-chip ZPL photon collection rate to the total saturated emission rate of the NV center. The saturated on-chip ZPL photon collection rate is obtained from the saturated off-chip count rate and the measured collection path efficiency $\eta_{cp}$ (see Section SI.1). The total saturated emission rate (including phonon sideband emission) is determined using the measured on-resonance lifetime of the selected NV center and a simple 5-level population density model~\cite{Robledo2011}. It is assumed that the NV centers are in the useful negatively charged state $\sim$70\% of the time under continuous wave 532 nm excitation~\cite{Aslam2013}. Using the measured on-resonance lifetime for Disk 1, a total saturated emission rate of $2.85\times10^7$ s$^{-1}$ is obtained. The estimated saturated on-chip collection rate of $2.48\times10^6$ s$^{-1}$ thus corresponds to $\eta_1=$ 9\%.

In the second method, we first use the measured Purcell factor to calculate the total quantum efficiency into the disk ($\eta_{disk}$) as:

\begin{equation}
	\eta_{disk} = \eta_{ZPL}\eta_{mode}
\end{equation}
\begin{equation}
	= \left(\frac{(F_P + 1)\Gamma_{ZPL}}{(F_P + 1)\Gamma_{ZPL} + \Gamma_{PSB}}\right)\left(\frac{F_P}{F_P + 1}\right)
\end{equation}
\begin{equation}
	= \frac{F_P}{(F_P+\Gamma_0/\Gamma_{ZPL})},
\end{equation}
where $\eta_{ZPL}$ is the proportion of photons emitting at the ZPL wavelength, $\eta_{mode}$ is the proportion of ZPL photons emitting into the disk mode, $F_P$ is the Purcell factor and $\Gamma_0(\Gamma_{ZPL},\Gamma_{PSB})$ is the total radiative emission rate (ZPL emission rate, phonon-sideband emission rate). Transmission measurements are used to obtain the disk-to-waveguide out-coupling efficiency {$\eta_{out}$,

\begin{equation}
	\eta_{out} = \frac{Q}{Q_c},
\end{equation}
where Q is the measured quality factor. The coupling-limited quality factor Q$_c$ can be calculated numerically using:

\begin{equation}
	Q_c = Q\left(\frac{1-t}{1-\gamma t}\right), 
\end{equation}
\begin{equation}
	\frac{T_{res}}{T_0} = \frac{(t-\gamma)^2}{(1-\gamma t)^2},
\end{equation}
where $t$ ($\gamma$) is the field transmission coefficient through the coupling region (resonator mode round trip), and $T_{res}/T_0$ is the normalized transmission at the resonance-dip minimum. All 4 devices for which $\eta_{out}$ was calculated were determined to be under-coupled based on large-scale transmission measurements~\cite{ref:gould2016lsg}. The total quantum efficiency is calculated as the product $\eta_2=\eta_{disk}\eta_{out}$. In the case of Disk~1, $\eta_{disk}$ was calculated to be 46\% and $\eta_{out}$ was determined to be 20\%. The resulting estimated total quantum efficiency is $\eta_2=$ 9\%, in agreement with $\eta_1$. For some disks, there is a discrepancy between the two estimation methods. This can be attributed to uncertainty in both the NV center charge-state ratio and the lifetime measurement.

\bibliography{fu_lab_bib,mikeSSPD_bib,mikeCountRate_bib,mikeLifetime_bib,mikeSI1_bib,mikeSI2_bib,mikeSI3_bib,mikeSI4_bib}

\end{document}